%% file: main.tex
\documentclass[letterpaper]{article} 
\usepackage{aaai25}  
\usepackage{times}  
\usepackage{helvet}  
\usepackage{courier}  
\usepackage[hyphens]{url}  
\usepackage{graphicx} 
\usepackage{natbib}  
\usepackage{caption} 
\usepackage{algorithm}
\usepackage{algorithmic}

\usepackage{xcolor}

\usepackage{multirow}
\usepackage{array}

\usepackage[most]{tcolorbox}

\usepackage{makecell}

\usepackage{caption}

\usepackage{booktabs}

\usepackage{placeins}

\usepackage{newfloat}
\usepackage{listings}
\DeclareCaptionStyle{ruled}{labelfont=normalfont,labelsep=colon,strut=off} 
\floatstyle{ruled}
\newfloat{listing}{tb}{lst}{}
\floatname{listing}{Listing}
\begin{document}

\title{What Comes After Harm? Mapping Reparative Actions in AI through Justice Frameworks}

\author{
    Sijia Xiao\textsuperscript{\rm 1},
    Haodi Zou \textsuperscript{\rm 2},
    Alice Qian Zhang \textsuperscript{\rm 1},
    Deepak Kumar \textsuperscript{\rm 2},
    Hong Shen \textsuperscript{\rm 1},
    Jason Hong \textsuperscript{\rm 1},
    Motahhare Eslami \textsuperscript{\rm 1}
}
\affiliations{
    \textsuperscript{\rm 1}Human-Computer Interaction Institute, Carnegie Mellon University\\
    \textsuperscript{\rm 2}Department of Computer Science and Engineering, University of California, San Diego\\


    xiaosijia@cmu.edu
%
}

\maketitle

\begin{abstract}

As Artificial Intelligence (AI) systems are integrated into more aspects of society, they offer new capabilities but also cause a range of harms that are drawing increasing scrutiny. A large body of work in the Responsible AI community has focused on identifying and auditing these harms. However, much less is understood about what happens after harm occurs: what constitutes reparation, who initiates it, and how effective these reparations are.
In this paper, we develop a taxonomy of AI harm reparation based on a thematic analysis of real-world incidents. The taxonomy organizes reparative actions into four overarching goals: acknowledging harm, attributing responsibility, providing remedies, and enabling systemic change. 
We apply this framework to a dataset of 1,060 AI-related incidents, analyzing the prevalence of each action and the distribution of stakeholder involvement. Our findings show that reparation efforts are concentrated in early, symbolic stages, with limited actions toward accountability or structural reform. Drawing on theories of justice, we argue that existing responses fall short of delivering meaningful redress. This work contributes a foundation for advancing more accountable and reparative approaches to Responsible AI.
\end{abstract}

%
\input{comments}

\input{body/introduction}
\input{body/related_work}

\input{body/method}

\input{body/finding_taxonomy}
\input{body/finding_quant_analysis}

\input{body/discussion}

\bibliography{zotero-references, manual_references, alice_references}

\input{body/appendix}

\end{document}

%% file: comments.tex
\definecolor{MyDarkBlue}{rgb}{0,0.1,1}
\definecolor{MyDarkGreen}{rgb}{0.02,0.6,0.02}
\definecolor{MyDarkRed}{rgb}{0.8,0.02,0.02}
\definecolor{MyDarkOrange}{rgb}{0.40,0.2,0.02}
\definecolor{MyPurple}{RGB}{111,0,255}
\definecolor{MyRed}{rgb}{1.0,0.0,0.0}
\definecolor{MyGold}{rgb}{0.75,0.6,0.12}
\definecolor{MyDarkgray}{rgb}{0.66, 0.66, 0.66}

\newcommand{\sx}[1]{\textcolor{MyDarkGreen}{[Sijia: #1]}}
\newcommand{\hz}[1]{\textcolor{MyRed}{[Haodi: #1]}}
\newcommand{\az}[1]{\textcolor{MyDarkOrange}{[Alice: #1]}}
\newcommand{\me}[1]{\textcolor{MyDarkBlue}{[Motahhare: #1]}}
\newcommand{\jh}[1]{\textcolor{MyPurple}{[Jason: #1]}}
\newcommand{\hs}[1]{\textcolor{MyGold}{[Hong: #1]}}
\newcommand{\dk}[1]{\textcolor{MyDarkRed}{[Deepak: #1]}}

\newenvironment{thisnote}{\par\color{MyDarkBlue}}{\par}

%% file: body/introduction.tex
\section{Introduction}
When New York City introduced a law requiring employers to disclose their use of AI systems in hiring and to publish audit results~\cite{nyc_aedt_2021}, it was widely celebrated as a breakthrough in algorithmic accountability. Policymakers and researchers saw it as a promising intervention to address bias and discrimination and to drive organizational change~\cite{johnson_ai_accountability_2021}. Yet, the law stops short of requiring companies to act on audit findings, such as correcting disparate impact. Critics have pointed out that organizations can comply in form but not in substance by disclosing superficial results or by inserting a human reviewer--- workarounds that create the appearance of accountability without delivering meaningful reform~\cite{wright2024null}.

This example illustrates a gap between recognizing harm and taking responsibility for addressing it. While recent research in Responsible AI (RAI) has made substantial progress in identifying and characterizing harms~\cite{schelble_towards_2024, zhang2024dark, techtonic2023inescapable} and in enhancing the auditing process~\cite{mokander2023auditing, metaxa_auditing_2021}, we still know relatively little about what actions are taken \textit{after} harm is acknowledged. In particular, it remains unclear how often those actions contribute to meaningful redress for those impacted, which is critical for moving beyond the appearance of oversight to meaningful accountability.

In this paper, we focus on the notion of reparative action. Grounded in theories of justice, we view reparation not merely as a pragmatic fix, but as a normative response to wrongdoing~\cite{un_reparation_2005}. We draw from three justice frameworks widely applied in the context of social and technological harm: punitive, restorative, and transformative justice \cite{feinberg2019expressive, zehr_little_2015, nocella2011overview}. Using these frameworks, we define ``AI harm reparative action" as \textit{the action taken in response to harm caused by AI systems that participates in the negotiation of redress and justice}. In this view, reparative action is not a single event, but a range of practices that play a role in shaping how accountability and repair are understood and enacted.

To understand what reparative actions entail, we first examine how reparative actions are (or are not) carried out following documented incidents of AI harm. We use the AIAAIC dataset,\footnote{https://www.aiaaic.org/aiaaic-repository/user-guide} a public repository of documented AI harms, to systematically investigate the post-incident actions taken by corporations, regulators, affected users, and other stakeholders.  We ask two research questions: \textbf{RQ(1)} What types of reparative actions are taken following AI harm? Who is involved in those reparative actions, and how? \textbf{RQ(2)} How are these actions distributed across cases, and what patterns emerge in the involvement of different stakeholder groups?

To explore this, we first qualitatively analyze a purposefully sampled subset of incidents and develop a taxonomy of AI harm reparative actions. We then apply this taxonomy across 1060 cases in the AIAAIC dataset to examine the prevalence of these reparative actions and stakeholders. We organize the reparative actions into four goals: \textit{Acknowledgment, Attribution, Remedy, and Reform}. Each goal encompasses distinct actions, ranging from perpetrators' communication to law and policy change. We then apply the taxonomy to see the distribution of actions and stakeholders in the full dataset. 

Our analysis reveals a stark imbalance: most responses stay at an Acknowledgement and Attribution level, such as public statements or third-party audits, while significantly fewer involve Remedy and Reform. This gap reflects a broader accountability shortfall in how the AI ecosystem responds to harm.

We draw on theories of justice to interpret what meaningful reparation could look like. Punitive justice emphasizes holding perpetrators accountable through consequences; restorative justice focuses on the needs and voices of those harmed; and transformative justice seeks to address the structural conditions that enabled harm. Yet current practices often fall short of these ideals, lacking clear accountability, offering limited support to affected communities, and rarely confronting systemic causes. We also highlight the vital but often overlooked role of affected communities and civil society in initiating responses, and suggest how future reparation practices might engage them. \footnote{A full version of this paper (with appendix) is at \url{https://arxiv.org/abs/2506.05687}.}

%% file: body/related_work.tex
\section{Related Work}
In this section, we review three areas of research relevant to our study. First, we summarize how responsible AI research has approached harm identification and assessment. Second, we examine what is known about responses to AI harm. Third, we draw on justice frameworks to provide conceptual grounding for analyzing reparation, adapting these traditions to the context of AI harm.

\subsection{Identifying and Assessing AI Harm}
We follow a normative definition of AI harm as a setback to an individual or group’s interests caused by the design, deployment, or use of an AI system \cite{diberardino2024algorithmic}. Existing research largely emphasizes harm detection at the design and development stages and evaluation after AI is created ~\cite{raji2020closing}. Researchers and practitioners increasingly turn to AI auditing, which involves systematic evaluations of AI systems to assess compliance with ethical, legal, or organizational standards~\cite{mokander2023auditing, metaxa_auditing_2021, birhane2024ai}. A related technique is AI red-teaming which adopts a more adversarial perspective on stress-testing AI systems and simulation of bad actors or harmful use cases to reveal emergent or latent risks~\cite{ganguli2022red}. On the other hand, benchmarking tools such as UnsafeBench~\cite{qu2024unsafebench} and AgentHarm~\cite{andriushchenko2024agentharm} enable quantification of potential risks. 

While these approaches are increasingly applied to post-deployment systems~\cite{casper2024black, singh2025red}, the goal remains oriented toward harm prevention and risk detection. Few provide mechanisms to address harm once it has occurred or to involve affected parties in shaping the response. To address this gap, emerging research has proposed participatory methods that incorporate users into the auditing and red-teaming process~\cite{lam_end-user_2022, cabrera2021discovering, shen_everyday_2021, deng_understanding_2023, deng2025weaudit}. These approaches draw on users’ situated knowledge to uncover harms that technical evaluations may overlook. However, scholars have questioned whether these methods truly foster accountability or merely serve as corporate risk management strategies~\cite{gillespie2024ai, feffer_red-teaming_2024}. Our work complements this line of work on detection and evaluation by shifting focus to the aftermath of harm.
We examine how AI harm is acknowledged and repaired: who takes action, in what form, and whether these responses align with the needs of those affected. 

\subsection{What Happens after AI Harm is Surfaced?}

While much attention has focused on preventing AI harm, a growing body of research examines how individuals and communities respond after harm has occurred. Studies have documented acts of resistance and refusal by affected users and workers~\cite{velkova_algorithmic_2021, ganesh2022resistance}, as well as conditions that prompt these responses. For instance, Johnson et al. analyzed the abandonment of algorithms following public backlash~\cite{johnson_fall_2024}, while Ehsan et al. show how withdrawn systems can still erode institutional trust~\cite{ehsan_algorithmic_2022}. Others have explored how people believe AI harms should be addressed, such as public preferences for punishment or accountability~\cite{lima_blaming_2023}, or how grassroots actors organize for redress in the face of institutional power~\cite{devrio_building_2024}. 

This growing literature has shed light on varied responses to harm, but often centers on individual cases, specific actors, or limited forms of accountability~\cite{bogiatzis2024beyond}. Our study contributes a systematic analysis of over 1,000 AI harm incidents. Rather than treating responses as isolated events, we conceptualize reparation as a series of actions unfolding over time and involving multiple stakeholders. In addition, we interpret these responses through the lens of reparation and justice, examining how obligations to repair harm are acknowledged, resisted, or enacted in practice.


\subsection{Reparative Action in Justice Frameworks}

This paper draws on three major justice traditions—punitive, restorative, and transformative justice—to contextualize different approaches to reparation. \textit{Punitive justice} emphasizes holding perpetrators accountable through proportional punishment, such as incarceration, fines, or legal sanctions~\cite{feinberg2019expressive, foucault2023discipline, ashworth2021sentencing}. It is the dominant model in most legal systems and widely used across online and offline governance \cite{gillespie_custodians_2018}. In contrast, \textit{restorative justice} centers the needs of those harmed, aiming to repair relationships and rebuild trust through practices such as public acknowledgment, apology, dialogue, and community-based support~\cite{zehr_little_2015, pranis_little_2015}. It has been widely applied in criminal justice and increasingly studied in digital governance and online harm contexts~\cite{xiao_addressing_2023, xiao_random_2020}. \textit{Transformative justice} builds on these traditions by addressing the structural conditions that enable harm, such as racism, ableism, and economic injustice~\cite{morris2000stories, kaba_we_2021}. It expands the idea of reparation beyond individual or relational repair, advocating for systemic change through institutional reform, power redistribution, and community-led design~\cite{kaba_fumbling_2019}.

These frameworks offer complementary perspectives in how they define harm, assign responsibility, and envision repair. Punitive and restorative justice both focus on interpersonal accountability, but take different approaches: punitive justice emphasizes external consequences such as punishment and deterrence, while restorative justice emphasizes mutual recognition, healing, and support for those affected \cite{wenzel2008retributive}. In addition, while restorative and transformative justice both center the harmed party, they differ in whether the aim is to restore relationships or to transform institutions \cite{kim2021transformative}.

These theories of justice have been widely applied to discussions of harm reparation in offline contexts and the governance of online communities \cite{nocella2011overview, 10.1145/3613904.3642704}. In the context of harm caused by AI systems, however, the process of reparation remains underexplored, and little attention has been given to the normative frameworks needed to assess the adequacy of such responses. In this study, we use these justice frameworks not only to interpret reparative actions, but also to evaluate how these actions contribute to meaningful repair.

%% file: body/method.tex
\section{Methodology}
We based our analysis on the AIAAIC repository of AI incidents, an independent, public-interest resource that documents events and controversies involving AI, algorithms, and automation\footnote{https://www.aiaaic.org/}. Compared to similar resources (e.g., the AI Incident Database), AIAAIC offers greater breadth and detail, with over 1000 incidents that include summaries, analyses of causes and implications, and links to related news coverage. It has been maintained by global wide volunteers and has been widely used in AI harm research to examine the consequences of AI failures and to develop harm taxonomies~\cite{lee2024deepfakes, johnson_fall_2024}.

Our analysis of the dataset involved two main stages: (1) developing a taxonomy of reparative actions based on a purposive sample of incidents and guided by justice frameworks, (2) identifying broader patterns by analyzing each incident with large language models (LLMs).

\subsection{Developing a Taxonomy of AI Harm Reparation}
We used all 1,060 complete incident entries submitted to the AIAAIC repository before December~2024 where each entry contains a description of the incident, details of post-incident developments, and references to external reporting. To construct our taxonomy of reparative actions, we analyzed a purposively sampled subset of incidents, as detailed below.

\subsubsection{Data Familiarization}
Given the dataset’s scale and diversity, we began with a familiarization phase to understand the structure and content of reported responses. Three researchers independently reviewed each incident summary and, when needed, consulted linked news articles to clarify timelines and stakeholder involvement. During this phase, we manually recorded the post-incident actions described in the 1060 incidents and identified the stakeholders responsible for initiating them (see Appendix A for an example). 

\subsubsection{Selection of Taxonomy Development Dataset}
Our initial review revealed that post-incident responses in the AIAAIC repository varied significantly. While some incidents involved concrete follow-up measures, such as algorithmic audits, product recalls, or financial compensation, we noticed many incidents that concluded with general communications or vague promises of future investigation. As a result, directly drawing a random sample from the full dataset risked underrepresenting reparative strategies.

To address this, we defined the criterion of \textit{substantial reparative action} as those leading to tangible outcomes beyond verbal communication. These include issuing refunds, redesigning products, initiating consequential investigations, and implementing staffing or policy changes. Isolated public statements or investigations without follow-through were excluded.

Using this criterion, we applied purposive sampling~\cite{campbell2020purposive} to identify 671 relevant incidents. From these, we randomly selected 200 to form the \textit{taxonomy development dataset}. To ensure analytical saturation, we spot-coded 20 additional incidents (5\% of the remainder)~\cite{lee_deepfakes_2024}. No new action types emerged, indicating that the sample sufficiently captured the range of reparative responses.

\subsubsection{Developing the AI Harm Taxonomy}
To construct our taxonomy of AI harm reparation, we conducted a thematic analysis~\cite{creswell_research_2017} of the 200-incident taxonomy development dataset. While our analysis was primarily inductive, it was informed by concepts and language drawn from restorative, punitive, and transformative justice, which offered conceptual grounding for distinguishing types of reparation.

We began by identifying and coding specific post-incident actions and the stakeholders who initiated them. For example, corporate responses such as apologies or explanations were grouped under Perpetrators’ Communication, while regulatory fines and legal charges were coded as Repercussions. Stakeholders were grouped based on their functional roles, such as Regulators and Government Bodies, Media, or Affected Users. While our action labels and categories were shaped by justice discourse, \textit{we did not assign individual actions to particular justice frameworks}, as many actions could be interpreted through multiple lenses depending on context.

In the next phase, we organized the action categories based on their underlying goals and orientations as reflected in justice frameworks. This approach allowed us to identify both common justice processes such as acknowledgment, accountability, and redress, and the distinctive contributions of each framework such as the emphasis on support to affected communities in restorative justice or systemic change in transformative justice. Through this lens, we developed four overarching goals in reparation: \textit{Acknowledgment, Attribution, Remedy, and Reform}. We refined the taxonomy through iterative team discussions over several months, shaping both action categories and stakeholder groupings. The final taxonomy appears in Table~\ref{taxonomy} and is elaborated in the findings section.

\subsection{Analyzing Reparative Actions At Scale}
We next examine how these reparative actions are distributed in the AIAAIC incident database.  To address this, we applied the taxonomy developed in the previous stage to all 1060 incidents. Given the scale of the dataset, manual coding was impractical. Recent research has demonstrated the potential of large language models (LLMs) to support deductive qualitative coding at scale~\cite{xiao2023supporting, mun2024particip}. We used GPT-4 Turbo, accessed via the OpenAI API\footnote{https://platform.openai.com}, to assist with this multi-label classification task, in which the model identified the presence or absence of each predefined reparative action and stakeholder category within incident descriptions.

We scraped each incident’s summary and linked articles to form a text corpus, then applied a two-step prompting pipeline. First, the model extracted stakeholder-action pairs, mirroring the structure of our manual coding. Second, it assessed whether each action category was present, classified responsible stakeholders, and provided supporting evidence. During prompt development, we conducted multiple rounds of qualitative error analysis to identify common failure patterns and refine the prompt. An example of the final prompt is included in Appendix~C. To validate this approach, we compared model outputs to human annotations for a random sample of 200 incidents (20 per action category). Three coders labeled action and stakeholder categories, achieving 93\% inter-coder reliability for actions and 90\% for stakeholders. Using these annotations as ground truth, the LLM achieved 87\% accuracy on actions and 79\% on stakeholders, indicating substantial agreement. Based on this validation, we applied the LLM to code the remaining incidents. Summary statistics and prompt details are provided in Appendix~D.

\subsection{Limitations}
Our analysis draws on the AIAAIC repository, which aggregates incidents primarily from public sources such as news articles. As a result, the dataset is not a comprehensive or random sample of AI harm and may overrepresent high-profile cases while overlooking less visible or underreported incidents. Certain responses—such as private actions by affected individuals~\cite{devrio_building_2024} or undisclosed internal changes by organizations~\cite{lukpat_ai_employees_2024}—are less likely to appear in public reporting, potentially leading to an underestimation of some stakeholders' involvement.

To scale our analysis, we used large language models (LLMs). While we validated outputs against human-coded samples and carefully designed prompts, LLM performance is shaped by training data and prompt design~\cite{guo2024evaluating}. We treat LLM-assisted coding as a tool to identify broad patterns, not a substitute for human interpretation.

%% file: body/finding_taxonomy.tex

\section{A Taxonomy of AI Harm Reparative Actions}
\subsection{Overview: Justice-oriented Organization of Taxonomy}

We present a taxonomy of AI harm reparative actions (Table~\ref{taxonomy}), organized according to four overarching goals that reflect key orientations in justice theory: \textit{Acknowledgment}, \textit{Attribution}, \textit{Remedy}, and \textit{Reform}. Rather than assigning each action to a specific justice framework, we used concepts from punitive, restorative, and transformative justice to group actions based on their underlying reparative aims. \textit{Acknowledgment} and \textit{Attribution} are foundational steps shared across justice traditions: addressing harm begins with recognizing its occurrence and establishing responsibility \cite{un_reparation_2005}. \textit{Remedy} aligns with both punitive and restorative justice, though its emphasis is on redress and interpersonal accountability \cite{zehr_little_2015, duff2005punishment}. \textit{Reform}, on the other hand, reflects the commitments of transformative justice, which seeks to address systemic conditions and prevent future harm \cite{morris2000stories}.

Each goal encompasses distinct types of actions and the stakeholders who initiate them\footnote{Stakeholder group definitions are provided in Appendix B.}. While these goals reflect a general progression from surfacing harm to pursuing structural change, they are intended as categories rather than a prescriptive sequence. In practice, incidents may engage with only a subset of these goals or pursue them in a non-linear or overlapping manner. This justice-informed structure allows us to evaluate not only what actions occur but also how they reflect broader commitments to accountability and repair.

\begin{figure}[t]
\centering
\includegraphics[width=\columnwidth]{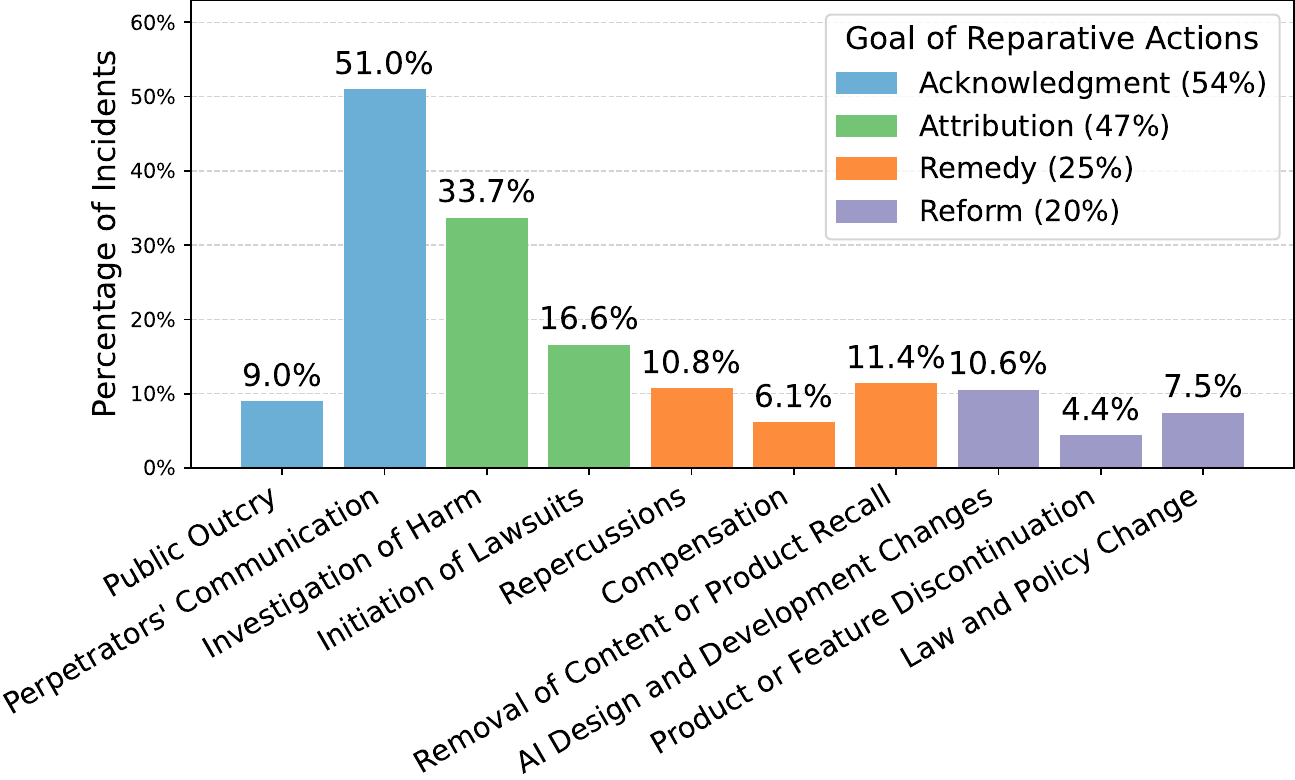}
\caption{\textbf{Distribution of Reparative Actions by Justice-Oriented Goals}. Perpetrators' communication was most common, while legal or policy change, compensation, and product discontinuation were among the least frequent, each appearing in fewer than 8\% of cases.}
\label{stage_and_actions}
\end{figure}

To contextualize the taxonomy and illustrate broader patterns, we provide a high-level summary of how frequently each justice-oriented goal appears across the dataset, offering a complementary quantitative view of the prevalence and distribution of reparative responses. In our analysis of 1,060 AI incidents, 54\% involved actions related to \textit{Acknowledgment} and 47\% involved \textit{Attribution}. By contrast, only 25\% of cases included any form of \textit{Remedy}, and 20\% showed evidence of \textit{Reform}. Figure~\ref{stage_and_actions} presents the distribution of specific actions across these goals. The most frequent action was communication from perpetrators, appearing in 541 cases (51\%). In comparison, more impactful forms of follow-through were rare: only 7\% of cases involved legal or policy reform, 6\% mentioned compensation, and just 4\% resulted in the discontinuation of a product or feature.

Below, we describe our taxonomy in detail. We refer to incidents using their AIAAIC case ID, abbreviated as A[case ID] (e.g., A1673).

\begin{table*}[t]
\centering
\small
\renewcommand{\arraystretch}{0.9}
\begin{tabular}{p{0.12\textwidth} p{0.25\textwidth} p{0.19\textwidth} p{0.33\textwidth}}
\toprule
\textbf{Goal} & \textbf{Reparative Action} & \textbf{Major Stakeholders} & \textbf{Examples} \\
\midrule

\multirow{2}{*}{Acknowledgment}
& Public Outcry & General Public, Advocacy Groups, Affected Users & Protests; Petitions; Online backlash \\
\cmidrule(l){2-4}
& Perpetrators' Communication & AI Corporations & Denials; Justifications; Apologies \\
\midrule

\multirow{2}{*}{Attribution}
& Investigation of Harm & Regulators, Media, \newline AI Corporations & Internal corporate audits; Government audits; Investigative journalism \\
\cmidrule(l){2-4}
& Initiation of Legal Actions & Affected Users, Advocacy Groups, Regulators & Pre-action letter; Initiation of lawsuits \\
\midrule

\multirow{3}{*}{Remedy}
& Repercussions & Regulators, \newline  AI Corporations & Fines; Criminal charges; Dismissals \\
\cmidrule(l){2-4}
& Compensation & AI Corporations, \newline End-Users & Refunds; Settlements \\
\cmidrule(l){2-4}
& Removal of Content \newline or Product Recall & AI Corporations &  Partial dataset removal; Product safety recall \\
\midrule

\multirow{3}{*}{Reform}
& Discontinuation & AI Corporations & Permanent product or service shutdown \\
\cmidrule(l){2-4}
& AI Design Changes & AI Corporations & Algorithm redesign; Adjustments to product safety protocols \\
\cmidrule(l){2-4}
& Law or Policy Change & Regulators, \newline AI Corporations & Public
legislation updates; Internal policy updates \\
\bottomrule
\end{tabular}
\caption{\textbf{Taxonomy of Reparative Actions.} This table presents our taxonomy of reparative actions, organized by four overarching goals: Acknowledgment, Attribution, Remedy, and Reform. For each goal, we list corresponding action types, the major stakeholders involved, and common forms these actions take.}
\label{taxonomy}
\end{table*}

\subsection{Acknowledgment}
Acknowledgment, the first step in responding to AI harm, seeks to make the harm visible and initiate collective sensemaking. While these actions do not provide material repair, they play a crucial role in shaping public narratives, surfacing concerns, and influencing how subsequent responses unfold. 

\subsubsection{Public Outcry}
We define public outcry as the expression of concern, anger, or dissatisfaction from members of the public, often emerging through online or grassroots channels. Public outcry occurred in 95 of the 1,060 cases (8.96\%). This action was primarily driven by the general public (70 cases), followed by affected users (28) and advocacy groups (16). These actors often acted as early signalers of harm, helping to surface incidents before formal accountability mechanisms were triggered.

One common form of public outcry is social media backlash. For example, when FaceApp released a ``hot'' filter that reinforced racist and Eurocentric beauty standards, users and critics voiced concern online, quickly drawing broader attention (A1673). Public outcry can also take more organized forms, including protests, petitions, crowdfunding efforts, or consumer boycotts. In one case, Fight for the Future launched a campaign against the University of British Columbia’s use of Proctorio, sparking widespread opposition to automated student surveillance. Their efforts were widely shared on social media and echoed by students and advocacy groups, catalyzing broader resistance to automated surveillance in education (A0473). The visibility and influence of such outcry often depend on amplification by journalists and advocacy groups, who help transform scattered responses into broader public narratives.


\subsubsection{Perpetrators' Communication}
We define perpetrators' communication as the public statements made by the party responsible for AI harm. This was the most frequent response in our dataset, appearing in 541 of 1,060 incidents (51.04\%), predominantly issued by AI corporations (489 cases) and end-users (76 cases). These statements are typically reactive, prompted by media scrutiny, legal threats, or public backlash. While they often appear reparative on the surface, they frequently serve reputational management goals rather than delivering accountability. Nonetheless, public communication can be an important first step in acknowledging harm and setting the stage for future action.

Such statements range from acknowledgments and promises of improvement to deflections of responsibility. When Air Canada's chatbot provided misinformation to a customer, the airline acknowledged the issue and promised to update the system to prevent recurrence. However, such acknowledgments are not always paired with substantive commitments (A1339). By contrast, when Tesla faced a lawsuit over a crash involving its autonomous driving system, it publicly released vehicle logs to suggest driver error, denying system fault (A0555). In another case, when the startup Moodbeam was criticized for emotional surveillance in the workplace, it defended the product as a tool for supporting remote workers, downplaying privacy concerns (A0515).

Despite their visibility, these statements do not often commit to remedy or structural reform. This may partly reflect practical constraints: systemic changes often require time, coordination, or factors beyond a single actor’s control. Still, the absence of meaningful follow-up creates a gap between symbolic acknowledgment and substantive repair. In many cases, these communications remain the only public-facing response, leaving affected communities without redress or closure.

\subsection{Attribution}
The second type of reparative action centers on assigning responsibility. These actions serve to legitimize harm claims, clarify accountability, and sometimes lay the groundwork for downstream reparative or punitive measures.

\subsubsection{Investigation of Harm}
We define investigations of harm as formal efforts to examine an AI incident. This action appeared in 357 incidents (33.68\%). Regulators and government bodies led most investigations (220 cases), followed by media organizations (77) and AI corporations (72). While all three groups play investigative roles, they differ in authority, transparency, and power to enforce consequences.

\textit{Regulatory investigations} carry the strongest mandate. Government and public agencies can issue fines, bans, or legal action. For example, Spain’s data protection authority fined Plastic Forte after finding it used facial recognition technology without proper consent (A1025).

\textit{Investigative journalism} plays a critical role in surfacing harm, especially when formal oversight is lacking. In one case, journalist Ko Narin exposed deepfake pornography in South Korean schools, prompting public outcry and eventually a government investigation into Telegram’s role in the abuse (A1727). However, journalistic investigations rarely lead to accountability on their own. For example, despite ProPublica’s exposé on McKinsey’s violence-associated analytics system at Rikers Island, the system remained in use (A0534).

\textit{Corporate-led audits} are internally driven and often lack transparency or external accountability. Figma’s CEO launched an internal review of an AI feature suspected of replicating copyrighted designs, but the audit’s findings were never fully disclosed (A1560).  Across such cases, we found that internal investigations are inconsistently reported and often lack transparency unless media or public scrutiny compels a response.

\subsubsection{Initiation of Legal Actions}
We define initiation of legal actions as the act of formally submitting a legal complaint in response to an AI harm incident. Initiation of Legal Actions appeared in 176 incidents (16.6\%). Some major stakeholder groups include affected users and non-users  (84 and 27 cases), advocacy groups (44), and regulators and government bodies (23). These cases are often protracted, with few reaching resolution during our study window.

\textit{Advocacy groups} frequently pursue legal accountability when other accountability channels fail. For instance, the UK-based Public Law Project issued a formal pre-action letter challenging the Home Office's algorithm for flagging ``sham marriages,'' citing discrimination and GDPR violations. As of the latest reporting, the case had not reached a resolution (A1389).

\textit{Affected Individuals, including public figures and professionals}, also initiated legal actions, typically over severe harm or personal rights violations. For example, several religious authors sued Meta and Microsoft for allegedly generating AI content derived from their copyrighted works (A1149). In another case, Megan Garcia filed a negligence lawsuit against Character AI after her son died by suicide following emotional attachment to a chatbot (A1781). Actress Scarlett Johansson also sued an AI developer for unauthorized use of her likeness (A1165).

\textit{Regulators and government bodies} also initiated legal proceedings, often with stronger enforcement power and clearer outcomes. For example, the U.S. Federal Trade Commission (FTC) sued Facebook for its role in the Cambridge Analytica scandal, resulting in a \$5 billion fine and a 20-year consent order. The UK Information Commissioner similarly imposed penalties and restrictions (A0128).

While legal actions provide a formal avenue for redress, our findings show they are typically slow-moving, legally complex, and more accessible to well-resourced individuals or organizations. Cases initiated by regulators are far more likely to result in concrete consequences than those led by private actors.

\subsection{Remedy}
These actions involves concrete responses aimed at addressing harm through punishment or material compensation. Unlike earlier actions that are largely symbolic or discursive, these actions impose consequences on responsible actors or offer redress to affected parties. 

\subsubsection{Repercussions} 
Repercussions are punitive consequences imposed on a responsible entity for AI harm. Repercussions were documented in 114~incidents (10.75\%). They were almost entirely enforced by regulators and government bodies (101 cases). These actions often arise when existing laws, such as those related to privacy, intellectual property, or safety, are violated in the deployment or use of AI systems.

Regulatory and legal authorities imposed a range of formal penalties, including financial fines, legal prosecution, and usage bans. Many of these actions targeted corporations for unlawful uses of AI technologies. For example, Sweden’s Privacy Protection Authority fined the national police for deploying Clearview AI’s facial recognition system without proper authorization in child abuse investigations (A0307). In another case, the U.S. Federal Trade Commission banned Rite Aid from using facial recognition technologies for five years and required new privacy and oversight measures (A1253). In rare but severe instances, individuals also faced criminal prosecution. In A1736, a Massachusetts man was arrested for cyberstalking after using AI tools to generate fake nude images and impersonate a victim through chatbot systems.

Some organizations also enacted internal consequences following public backlash or controversy. For instance, the editor-in-chief of Die Aktuelle, a German magazine, was fired after publishing an AI-generated fake interview (A0995). While these actions represent a form of accountability, they are not the result of formal legal enforcement and are typically not subject to public oversight or consistent standards.

Taken together, these cases show that while punitive consequences are possible, they remain relatively limited in scope and frequency. Most occur within the boundaries of existing legal frameworks and require institutional authority to be meaningfully enforced, highlighting the challenges of ensuring accountability without regulatory mechanisms.

\subsubsection{Compensation}
We refer to compensation as monetary or material remedies provided to individuals or groups who experienced harm in an act to address the damage done. This distinguishes it from financial penalties discussed under Repercussions, which are generally paid to regulatory bodies rather than harmed parties.  Compensation occurred in 65 incidents (6.13\%), making it one of the least frequent actions. AI corporations were the main providers of compensation (53 cases).

Compensation occurred both through legal processes and voluntary corporate action. In some cases, companies provided compensation as part of formal legal proceedings or in response to legal pressure. For example, DoorDash paid \$2.5 million to resolve a class action lawsuit over misleading tipping practices (A0537), and a Milan court ordered Google to pay €3,800 in damages to an entrepreneur for reputational harm caused by autocomplete suggestions (A1087). In other cases, compensation was offered without a public legal mandate. General Information Services and its subsidiary e-Background-checks.com provided \$10.5 million in relief to customers affected by inaccurate background checks, though the case did not result in a court ruling (A0853).

Among all reparative actions, compensation is the only one that provides direct material remedy to those affected by AI harm. Yet even in these cases, affected communities often have limited agency over whether compensation occurs, how it is structured, or who receives it. Instead, decisions about compensation are typically made through corporate discretion or legal negotiation, with little involvement from those directly impacted.

\subsubsection{Removal of Content or Product Recall}
This action refers to the removal of harmful AI-generated content or the recall of AI products to prevent further use or distribution. It was documented in 121 incidents (11.42\%). AI corporations led most of these actions (97 cases), followed by regulators that apply AI products (20) and end-users (16).

Content and product removals typically aim to eliminate specific sources of harm while preserving the broader AI system. Companies removed training data containing personal images of Australian children after it was revealed that the content had been collected without consent  (A1569).  Tesla recalled approximately 50,000 vehicles and discontinued its ``Assertive'' driving mode following regulatory concerns about safety compliance (A0816).

These actions play an important role in halting ongoing harm and responding to public concern. By removing harmful content or disabling risky features, they can offer immediate relief and demonstrate responsiveness from companies or regulators. However, such actions are typically reactive and limited in scope. Because they do not address underlying design flaws or systemic oversight gaps, similar harms may recur in new contexts. 

\subsection{Reform}
The final set of reparative actions involves structural changes intended to prevent future harm. These responses go beyond addressing individual incidents and aim to reshape the broader systems, products, or legal frameworks governing AI. While reform holds the greatest potential for long-term accountability and systemic improvement, our analysis shows that such efforts are inconsistently applied and often lack enforceable mechanisms to ensure follow-through.

\subsubsection{Product and Feature Discontinuation}
The action refers to the permanent or long-term discontinuation of an AI product, a specific feature, or an associated dataset. Unlike removal of content or features, which targets specific outputs, discontinuation reflects a decision to retire the system itself. It was the least frequent action in our dataset, occurring in 47 of the 1060 incidents (4.43\%), primarily carried out by AI corporations (39 cases).

Notable examples include Microsoft’s decision to permanently shut down its chatbot Tay after repeated instances of offensive content, despite multiple attempts at remediation (A045). Similarly, FaceApp discontinued its ``hot filter'' following public criticism over its reinforcement of racist and Eurocentric beauty standards (A1674).

Product or feature discontinuation can be a meaningful response when companies withdraw technologies that cause harm: it signals that certain systems that caused harm are no longer acceptable. However, companies are not obligated to explain or sustain these decisions, and it is often unclear whether discontinued features remain permanently retired. In some cases, similar functionalities may reappear in modified forms, suggesting the need for greater transparency and follow-through in such reforms.

\subsubsection{AI Design and Development Changes}
We define AI and design changes as technical changes made to the AI system to address the underlying issue and prevent recurrence. AI Design and Development Changes were present in 112 incidents (10.57\%). These were nearly all initiated by AI corporations (109 cases).  These actions aim to prevent recurrence by addressing the design and implementation issues that enable harm to happen.

Examples of this action include both changes to AI models and the implementation of safeguards in development and deployment processes. LinkedIn revised its name prediction algorithm to reduce gender bias by incorporating more diverse name datasets (A044). In another case, following accusations that its AI-generated video technology was being used to misrepresent or exploit actors, Synthesia updated its content development pipeline (A1787).

These changes reflect a proactive orientation toward harm prevention and growing attention to responsible design. Like other reform-oriented actions in our taxonomy, they are typically initiated at the discretion of companies and implemented without external oversight. This raises broader questions about how responsibility is defined and enacted across the AI industry.

\subsubsection{Law and Policy Changes}
We define this action as the introduction or revision of legal frameworks, regulations, or organizational policies in response to the AI incident. Law and Policy Change was observed in 79 incidents (7.45\%). Most were led by regulators and government bodies (54 cases), with AI corporations contributing to 29 cases.

These changes fell into two primary categories: public legislation and internal policy updates. On the legal side, governments introduced or amended laws to directly regulate AI-related harms. For example, Taiwan’s National Legislature amended its Criminal Code to criminalize the creation and distribution of deepfakes (A0771), and South Korea passed a law banning the possession or viewing of sexually exploitative deepfakes (A1727). Organizational policies were also revised in response to specific incidents. TikTok updated its privacy policy to include a Dutch-language version following user complaints (A0429). In another case, the AI video platform Synthesia implemented stricter content controls after a defamatory incident, restricting content creation to verified enterprise users and banning topics such as politics and race (A1093).

These reforms represent some of the most concrete efforts to govern AI systems and mitigate harm at scale. They demonstrate increasing awareness among both governments and organizations of the need for enforceable standards and structural safeguards. One important pattern is the geographic diversity of these responses, which reflects the global reach of AI but also highlights the fragmented nature of its regulation. In the absence of coordinated standards, policy changes tend to emerge in isolated jurisdictions, resulting in uneven protections and regulatory gaps across regions.


%% file: body/finding_quant_analysis.tex
\section{Stakeholder Participation Across Reparative Actions}
While our taxonomy identifies which stakeholder groups initiate each type of reparative action, it is organized around actions and does not reveal how individual stakeholders contribute across the broader set of actions. To complement this, Table~\ref{stakeholder_distribution} presents the frequency with which each stakeholder group initiates each action, with percentages reflecting their occurrence across all 1,060 incidents. This comparative view highlights patterns of responsibility and exposes disparities in who drives different forms of reparation.


\begin{table*}[t]
\centering
\small
\renewcommand{\arraystretch}{1.3}
\begin{tabular}{|>{\footnotesize}p{3.5cm}cccccccccc|}
\hline
\textbf{Stakeholder Group} &
\textbf{Outcry} &
\textbf{Comm.} &
\textbf{Invest.} &
\textbf{Legal.} &
\textbf{Repr.} &
\textbf{Comp.} &
\textbf{Recall} &  
\textbf{Design} &  
\textbf{Discont.} &
\textbf{Policy} \\
\hline
\makecell[l]{Perpetrators --\\AI Corporations} & -- & \textbf{46.1\%} & 6.8\% & 0.9\% & 1.1\% & 5.0\% & 9.2\% & 10.3\% & 3.7\% & 2.7\% \\
\hline
\makecell[l]{Perpetrators --\\End-Users} & 0.1\% & \textbf{7.2\%} & 0.7\% & 0.4\% & 0.3\% & 0.9\% & 1.5\% & 0.1\% & 0.5\% & 0.4\% \\
\hline
Affected Users & 2.6\% & -- & 0.1\% & \textbf{7.9\%} & -- & 0.1\% & -- & -- & -- & -- \\
\hline
\makecell[l]{Affected\\Non-Users} & 0.6\% & -- & -- & \textbf{2.6\%} & -- & -- & -- & -- & -- & -- \\
\hline
General Public & \textbf{6.6\%} & -- & -- & -- & -- & -- & -- & -- & -- & -- \\
\hline
\makecell[l]{Regulators and\\Government Bodies} & 0.2\% & 0.3\% & \textbf{20.8\%} & 2.2\% & 9.5\% & 1.2\% & 1.9\% & 0.8\% & 0.6\% & 5.1\% \\
\hline
Academia & 0.1\% & -- & \textbf{0.6\%} & -- & -- & -- & -- & -- & -- & -- \\
\hline
Media & 0.3\% & -- & \textbf{7.3\%} & 0.5\% & -- & -- & 0.2\% & -- & -- & -- \\
\hline
\makecell[l]{Auditors and\\Oversight Boards} & -- & -- & \textbf{0.8\%} & -- & -- & -- & -- & -- & -- & -- \\
\hline
Advocacy Groups & 1.5\% & -- & 1.0\% & \textbf{4.2\%} & -- & 0.1\% & -- & -- & -- & -- \\
\hline
\end{tabular}
\caption{\textbf{Stakeholder involvement across reparative actions.} Percentages represent the proportion of all 1,060 incidents. Dashes indicate no involvement; bolded values show each stakeholder’s highest level of engagement. AI corporations focused on communication, while structural remedies saw limited participation. Affected communities engaged infrequently; third parties surfaced harm but rarely shaped outcomes.  Please refer to Table~\ref{taxonomy} for complete action names.}
\label{stakeholder_distribution}
\end{table*}

\textit{Perpetrators primarily respond through communication}. AI corporations, the primary actors responsible for harm in the dataset, were involved in public communication far more often than in any other form of response. Their involvement in statements or apologies was more than four times higher than their participation in actions like compensation, product recalls, or design changes. This pattern suggests that perpetrators are more likely to manage perception than to initiate structural or corrective measures.

\textit{Affected communities have limited influence on outcomes.} Affected users most commonly appeared in public outcry or the initiation of lawsuits, typically in reactive roles. Affected non-users were even less visible, with limited involvement primarily in legal action. Both groups were largely absent from compensation, design changes, or policy reform, indicating limited influence over how harm is addressed.

\textit{Third parties help surface harm, but rarely guide resolution.} Media and academia had significant but secondary roles. Journalists were involved in 77 investigations, and academics contributed in more limited cases. Advocacy groups, while not directly involved in harm, play a part in public outcry, investigation of harm, and initiation of lawsuits. The general public also plays a significant role in engaging in public discourse about the harm. 

\textit{There is a lack of multi-stakeholder collaboration.}
Finally, we examined the extent of coordination across different stakeholder types. Our further analysis showed that only 304 incidents (29\%) involved any action carried out by more than one stakeholder group. The most common multi-actor responses were investigations (127 incidents), perpetrators’ communication (55), and lawsuits (65). In most cases, however, actions were isolated efforts by individual groups, suggesting a fragmented and uncoordinated approach to AI harm reparation.



%% file: body/discussion.tex
\section{Discussion}
This section examines how post-incident reparation efforts reflect, or fall short of, the pursuit of justice in AI harm. We begin by analyzing how actions aimed at remedy and reform correspond to theories of punitive, restorative, and transformative justice. While many of these actions seek to address harm, they often lack clear accountability, provide limited reparation to affected communities, or fail to engage with the structural conditions that produced the harm. We then turn to participation in the reparation process, focusing on how affected communities and third-party actors contribute to reparation, though often through informal or constrained roles.

\subsection{A Vacuum of Justice: Evaluating Reparative Actions Through Justice Frameworks}
Justice frameworks not only help categorize reparative actions but also reveal how current responses fall short of meaningful accountability and redress. Rather than offering prescriptions, these frameworks serve as interpretive tools for identifying limitations and possibilities. This approach aligns with calls in Responsible AI to develop context-sensitive, relational models of accountability~\cite{metcalf2023taking}. Below, we examine how actions under \textit{Remedy} reflect principles of punitive and restorative justice, and how those under \textit{Reform} correspond to transformative justice.

\subsubsection{Remedy for Past Harm: Rare Accountability and Exclusion of the Harmed}
Corrective justice traditions emphasize that those responsible for harm should be held accountable and that concrete efforts should be made to repair the losses experienced by those affected~\cite{feinberg2019expressive, zehr_little_2015}. These principles underlie both punitive and restorative justice, which guided our analysis of actions under the goal of \textit{Remedy}.

\textit{Punitive justice} holds that wrongdoers should face consequences proportionate to the harm they caused~\cite{feinberg2019expressive}. In our dataset, however, punitive responses were rare. Only 10\% of cases involved fines, bans, or dismissals, typically enforced by regulators or courts. We found much less evidence of internal accountability, such as companies publicly disclosing disciplinary action or self-imposed sanctions. While internal measures may occur, the lack of transparency makes them difficult to assess. This aligns with prior research showing that AI professionals often view accountability as diffuse and located outside their individual or organizational control~\cite{lancaster2024s}. Moreover, when punitive actions take place, they generally relied on traditional legal frameworks, such as privacy or copyright violations, rather than mechanisms tailored to AI-specific harms. This limits the effectiveness of punitive measures in addressing the unique and evolving risks of AI systems.

\textit{Restorative justice} focuses on repairing harm by centering the needs and agency of those affected and by enabling acknowledgment, dialogue, and restitution~\cite{zehr_little_2015}. Yet restorative practices were similarly scarce. Only 7\% of cases involved direct remedy to affected communities in the form of compensation. Few included meaningful input from those harmed in shaping the response. While public apologies and statements of intent were more common, they were often vague, symbolic, and lacked follow-through.  As prior work has noted, such gestures may function more as reputational risk management than as genuine accountability~\cite{metcalf2019owning, green2021contestation}. These patterns reflect a broader gap in current practices: even when harm is acknowledged, efforts to rebuild trust or restore relationships are limited. Despite growing calls for participatory, victim-centered approaches~\cite{ajmani2024data, ganesh2022resistance, velkova_algorithmic_2021}, most responses failed to return agency to affected communities, leaving the core aims of restorative justice unmet.

\subsubsection{Reform for Structural Harm: Structural Responses Without Structural Change}

Actions under the goal of \textit{Reform} aimed to address institutional or systemic conditions, such as organizational policies or algorithm designs. These responses align with the aims of \textit{transformative justice}, which seeks to prevent future harm by confronting and altering the structural and social conditions that allow harm to persist~\cite{morris2000stories}. However, in practice, such actions often fell short of these goals. 

In our dataset, 20\% of cases involved actions that could be interpreted as targeting structural factors. These included product and feature discontinuation, system redesigns, and law and policy change. However, many of these interventions were narrow in scope. They rarely addressed upstream drivers such as exploitative data practices, profit incentives that reward harmful behavior, or recurring harms across platforms. Most changes were confined to a single organization and lacked coordination across the broader ecosystem. 
In some cases, structural change took the form of product discontinuation rather than substantive redesign---what has been critiqued as a superficial fix that evades accountability~\cite{johnson_fall_2024}. These patterns echo broader critiques that ethics initiatives in the tech industry often serve to preempt regulation and manage public image rather than drive systemic accountability~\cite{metcalf2019owning, green2021contestation}.

\subsubsection{Remedy and Reform as Complementary Forms of Reparation}
A key distinction in our taxonomy is between reparative actions that address past harm (Remedy) and those that restructure systems to reduce recurrence (Reform). This distinction helps clarify a gap noted in existing scholarship: systemic changes cannot replace direct redress to those harmed. Recent critiques have pointed out that widely used accountability practices—such as auditing and red teaming—often mitigate risk rather than ensure meaningful reparation~\cite{gillespie2024ai, feffer_red-teaming_2024}. Similarly, AI policy frameworks frequently prioritize transparency and procedural oversight while lacking enforceable requirements for remediation~\cite{wright2024null}. Centering reparation requires that both individual and systemic dimensions be addressed. Remedy and reform must work together to ensure harms are not only acknowledged but actively repaired through inclusive and sustained action.

\subsection{Who Gets to Repair? Participation in AI Reparation}
Beyond mapping past responses, our taxonomy offers a forward-looking tool for guiding more accountable and participatory approaches to AI harm reparation. For affected communities, it provides language to articulate what they are owed and where redress is lacking. For companies and policymakers, it highlights neglected levers for remedy and reform. Researchers and auditors may use it to assess the distribution of responsibility, benchmark institutional responses, and track progress over time.

Yet while our taxonomy outlines a range of potential contributions, carrying out these roles in practice often faces structural and institutional barriers, as seen across our AI incident analysis. These challenges limit the ability of stakeholders, particularly those not involved in AI development or regulatory processes, to meaningfully engage in reparation. Below, we highlight two groups often marginalized in reparative processes: affected communities and third-party actors.


\subsubsection{Constrained Agency: Barriers to Collective Redress for Affected Communities}
Although affected communities are central to the moral case for reparation, they are often sidelined in shaping responses. Our analysis shows that they are rarely involved in formal processes of remedy or reform. For example, only 7\% of cases in our dataset involved compensation, and almost none included evidence of participatory design or consultation with affected individuals.

Moreover, it is difficult for affected individuals to mobilize collectively, particularly when harms are diffuse, individualized, or legally ambiguous. Legal actions, one of the few formal channels for redress, were more often initiated by regulators or advocacy groups than by those directly affected. This reflects deeper structural barriers: affected communities often lack the resources, visibility, or institutional support to demand meaningful accountability. Even when users engage in public contestation of AI harm, these efforts rarely connect with formal mechanisms of redress~\cite{velkova_algorithmic_2021, ganesh2022resistance}. Addressing this gap requires rethinking reparation as a participatory process that centers the agency and priorities of those harmed. Without such inclusion, reparation risks reinforcing the very power asymmetries it aims to challenge.

\subsubsection{Reparation from the Outside In: The Role of Public and Civil Society Actors in Reparation}

Our analysis highlights a broader ecosystem of actors who influence reparation and accountability, especially where formal mechanisms are lacking. While multi-stakeholder AI governance research often centers on developers and directly affected users~\cite{morley2020initial, rakova_where_2020, mohamed2020decolonial}, we extend this view to include civil society and the public, including advocacy groups, journalists, academics, and community members.

Prior Responsible AI scholarship has documented what these actors do to resist or prevent harm, such as advocacy groups mobilizing campaigns~\cite{contreras2023civilrightsai}, journalists framing incidents as systemic issues~\cite{angwin2016machine}, academics producing audits and critiques~\cite{deng2025weaudit, shen_everyday_2021, cabrera2021discovering}, and the public amplifying these efforts through petitions and media engagement~\cite{devrio_building_2024}. Building on this work, our contribution is to place these activities within the reparation process and show how they operate as oversight and public pressure that can create pathways toward remedy and reform.

Yet despite their impact, these interventions remain loosely connected to formal mechanisms such as legal remedies, compensation, or policy reform. Prior work notes that civil society’s role is often fragmented and reactive~\cite{raji2022outsider, lucaj2023ai, gillespie2024ai}, reflecting structural barriers: these actors are not directly affected by every incident and lack formal standing in governance. Still, their participation is essential given the societal reach of AI harms. Institutions can bridge this gap by creating channels for collaboration—supporting independent audits, consulting civil society in policy design, and enabling sustained involvement in reparative processes~\cite{lam_end-user_2022, feffer_red-teaming_2024}. Better integration can transform public pressure into institutional change and align reparation with the broader public interest.

\section{Conclusion}
This paper introduces a taxonomy of AI harm reparative actions and applies it to a large-scale analysis of post-incident responses. The taxonomy traces how harm is acknowledged, addressed, remedied, and, in some cases, transformed over time. Our findings reveal the uneven distribution of reparative efforts and the limits of current responses in achieving meaningful redress. We argue for more systemic approaches that center affected communities, integrate third-party oversight, and tie commitments to enforceable accountability.

%% file: body/appendix.tex
\clearpage
\onecolumn
\appendix
\section*{Appendices}

\section{Appendix A: Data Familiarization Example}
During data analysis, we began with a familiarization phase to understand the structure and content of reported responses. For all 1060 cases, we manually recorded the post-incident actions and identified the stakeholders responsible for initiating them. For example, for the incident involving TikTok's mishandling of minors' personal data in the Netherlands (incident ID AIAAIC0429), the recorded sequence of actions and stakeholders was: (1) Dutch Data Protection Authority (DPA): Investigated TikTok's handling of minors' data. (2) Consumentenbond (Dutch Consumer Organization): Filed €1.5B claim over unlawful data collection. (3) Dutch DPA: Fined TikTok €750K for lacking Dutch privacy notice. (4) TikTok/ByteDance: Updated its privacy policy to include a Dutch version and tightened teen safety settings.

\section{Appendix B: Stakeholder Definitions}
\renewcommand{\arraystretch}{1.2}
\begin{table}[H]
\centering
\caption{Definitions of primary stakeholder categories referenced in our taxonomy.}

\begin{tabular}{|p{0.28\textwidth}|p{0.62\textwidth}|}
\hline
\textbf{Stakeholder} & \textbf{Definition} \\
\hline
Perpetrators – AI Corporations & Companies that develop, deploy, or commercialize AI systems that caused harm. \\
\hline
Perpetrators – End-Users & Individuals or organizations that used AI systems and caused harm through their use. \\
\hline
Affected Users & People or groups who used the AI system as intended but were harmed. \\
\hline
Affected Non-Users & People or groups who did not use the AI system but were still harmed. \\
\hline
General Public & Members of society not directly harmed but who express concern or demand change. \\
\hline
Regulators and Government Bodies & Courts, regulators, and agencies that investigate or impose consequences. \\
\hline
Academia & Researchers or institutions who analyze incidents or propose solutions. \\
\hline
Media & Journalists and news outlets reporting on the incident. \\
\hline
Auditors and Oversight Boards & Entities that assess AI risks or harms through audits or review. \\
\hline
Advocacy Groups & NGOs or civil society organizations that represent affected communities. \\
\hline
\end{tabular}
\label{stakeholders}
\end{table}

\section{Appendix C: LLM prompt example}

\noindent
\begin{minipage}{\textwidth}
\setlength{\parskip}{0.5em}  

We use the Public Outcry category as an example to illustrate the LLM prompt we provided. The text input consists of a full
list of actions and stakeholders extracted from the case summary and related news articles in the AIAAIC database. Please note
that these represent preliminary extraction categories used during data collection and are distinct from the refined taxonomy
presented in our findings.

\textbf{Step 1:} Determine whether the text explicitly mentions any \textit{Public Outcry}.

\textbf{Definition:} Public Outcry refers to expressions of dissatisfaction, concern, or anger voiced publicly in response to an AI harm incident.

\textbf{Valid examples (for reference only):}
\begin{itemize}
  \item Social media backlash
  \item Protests and walkouts
  \item Public petitions
\end{itemize}

\textbf{Additional Rules:}
\begin{enumerate}
  \item Do not include private or internal complaints. Only include actions that reflect broad public visibility and concern.
  \item Reports by news media do not qualify as public outcry unless they describe public reactions (e.g., protests, petitions, social media backlash).
  \item Do not infer or assume – only include what is explicitly stated.
\end{enumerate}

\textbf{Step 2 (if Step 1 is positive):} Identify Initiating Stakeholders

\textbf{Stakeholder Categories:} \textit{[Insert stakeholder categories presented in Table~\ref{stakeholders}]}

\textbf{Output Format:}
\begin{itemize}
  \item If no public outcry is explicitly mentioned, output: \texttt{N/A}
  \item If yes, list each initiating stakeholder in the format: \\
  \texttt{[Stakeholder Category] -- [Name from text] -- [Direct evidence from text]}
\end{itemize}

\textbf{Do not include any additional explanation.}

\textbf{Text for analysis:} \textit{[Insert case summary and relevant news article text here]}

\end{minipage}

\section{Appendix D: LLM accuracy and Inter-Rater Reliability}

\subsubsection{Inter-rater reliability}
We assessed inter-rater reliability among three coders, each of whom independently coded five shared cases with every other coder. This resulted in six pairwise comparisons across the three coders. For accuracy of reparative action category, we calculated the percentage of actions where both coders agreed on whether an action was present. The average agreement across all pairs was 93.33\%. For accuracy of stakeholder category, we assigned a score of 1 for full agreement, 0.5 for partial agreement, and 0 for disagreement. These scores were averaged across five cases per pair. The average stakeholder agreement was 90.00\%.

\subsubsection{LLM accuracy}
To validate the accuracy of GPT-assisted coding across 1,060 incidents in the AIAAIC database, we conducted human annotation on a stratified sample of 200 cases—20 cases for each of the 10 reparative action categories in our taxonomy. For each category, we selected a balanced mix of positive and negative examples (10 each) and evaluated two components: whether the action was correctly identified, and whether the initiating stakeholders aligned with human labels. \textit{Action accuracy} was calculated as the proportion of cases where GPT's binary label (present or not) matched the human-coded label, out of 20 total cases per category. \textit{Stakeholder accuracy} was evaluated only for cases where the action was labeled as present by human annotators. In those cases, we compared the set of initiating stakeholders labeled by GPT against human annotations, assigning a score of 1 for exact matches, 0.5 for partial overlap, and 0 for no overlap. We report both macro-averaged scores across the ten action categories and micro-averaged scores weighted by the number of human-positive cases. Table~\ref{gpt_accuracy} reports the percent agreement for each action and stakeholder category, and the total accuracy.

\begin{table}[h]
\centering
\caption{LLM Classification Accuracy}
\label{gpt_accuracy}
\begin{tabular}{lcc}
\toprule
\textbf{Reparative Action Category} & \textbf{Action Accuracy} & \textbf{Stakeholder Accuracy} \\
\midrule
Public Outcry & 85.00\% & 77.27\% \\
Perpetrator’s Communication & 85.00\% & 81.82\% \\
Investigation of Harm & 80.00\% & 65.00\% \\
Initiation of Legal Actions & 95.00\% & 66.67\% \\
Repercussions & 95.00\% & 88.89\% \\
Compensation & 80.00\% & 75.00\% \\
AI Design and Development Change & 80.00\% & 81.25\% \\
Removal of Content or Product Recall & 90.00\% & 65.00\% \\
Product or Feature Discontinuation & 85.00\% & 92.86\% \\
Law and Policy Change & 95.00\% & 88.89\% \\
\midrule
\textbf{Overall} & \textbf{87.00\%} & \textbf{78.78\%} \\
\bottomrule
\end{tabular}
\end{table}